\documentclass[pdflatex,sn-mathphys-ay]{sn-jnl}

\usepackage{graphicx}%
\usepackage{multirow}%
\usepackage{amsmath,amssymb,amsfonts}%
\usepackage{amsthm}%
\usepackage{mathrsfs}%
\usepackage[title]{appendix}%
\usepackage{xcolor}%
\usepackage{textcomp}%
\usepackage{manyfoot}%
\usepackage{booktabs}%
\usepackage{algorithm}%
\usepackage{algorithmicx}%
\usepackage{algpseudocode}%
\usepackage{listings}
\usepackage{subcaption}

\theoremstyle{thmstyleone}%
\theoremstyle{thmstyletwo}%

\theoremstyle{thmstylethree}%

\begin{document}

\title[Generalized Bayesian Ensemble Survival Tree model]{Generalized Bayesian Ensemble Survival Tree (GBEST) model}


\author*[1]{\fnm{Elena} \sur{Ballante}}\email{elena.ballante@unipv.it}

\author[2]{\fnm{Pietro} \sur{Muliere}}

\author[1]{\fnm{Silvia} \sur{Figini}}

\affil*[1]{\orgdiv{Department of Political and Social Sciences}, \orgname{University of Pavia}, \orgaddress{\city{Pavia}, \country{Italy}}}

\affil[2]{\orgdiv{Department of Decision Sciences}, \orgname{Bocconi University},\orgaddress{ \city{Milano},  \country{Italy}}}


\abstract{This paper proposes a new class of predictive models for survival analysis called Generalized Bayesian Ensemble Survival Tree (GBEST). 
It is well known that survival analysis poses many different challenges, in particular when applied to small data or censorship mechanism. 
Our contribution is the proposal of an ensemble approach that uses Bayesian bootstrap and beta Stacy bootstrap methods to improve the outcome in survival application with a special focus on small datasets.
More precisely, a novel approach to integrate Beta Stacy Bayesian bootstrap in bagging tree models for censored data is proposed in this paper.
Empirical evidence achieved on simulated and real data underlines that our approach performs better in terms of predictive performances and stability of the results compared with classical survival models available in the literature.\\
In terms of methodology our novel contribution considers the adaptation of recent Bayesian ensemble approaches to survival data, providing a new model called Generalized Bayesian Ensemble Survival Tree (GBEST). A further result in terms of computational novelty is the implementation in R of GBEST, available in a public GitHub repository.}

\keywords{Survival analysis, Bootstrap, Bayesian non-parametric learning, ensemble models}



\maketitle

\section{Introduction}\label{sec1}
Survival analysis is a common framework in different application areas, as medicine, engineering, economy and many others. It is involved every time a phenomenon is observed until an event of interest occurs, and the final aim is to make inference or prediction on the time of the event. Historically the event is the death of patients (as in \cite{death}) but there are many other examples as relapse (see \cite{relapse}), and non medical examples as graduation (\cite {graduation}), student dropout (\cite{dropouts}), costumer churn (\cite{churn_mobile}), failure of a machine (\cite{failure}), credit default (\cite{GBST, default}) and so on.  
In order to tackle the variety of problems that emerge in these fields, the need of different methods and models that provide reliable results is felt by statisticians and data scientists. This need is specifically referred to the most difficult cases in this kind of analysis, for example when the sample size is extremely small (less than 100 statistical units) or the censorship level is high (more than 20/30\% of censored observations).  \\
In the context of non-parametric modelling, survival trees, defined in \cite{survtrees}, are a popular approach based on recursive partitioning of the data space. They are vastly used for the simplicity of interpretation of the results, as it is very easy to identify the important variables and the also provide a threshold for numerical covariates used. 
But survival tree models suffer of high variance, like decision trees defined in classification and regression framework 
(\cite{breiman}). In fact, they are very unstable, and a little variation in the training set can lead to a completely different tree. Even if the predictive power can still be good, this mine the reliability of the results in terms of interpretation. \\
To tackle the problem of instability and to improve the predictive performances, a classical procedure is to implement an ensemble model, i.e. a method to aggregate different models with the idea that the prediction obtained by multiple models will be better of the prediction of the single ones.\\
The most common approaches in ensemble models relies on the method of bagging as defined in \cite{bagging_breiman}, together with the boosting method that is outside the scope of this paper.\\
In survival analysis different bagging approaches are proposed, as bagging survival trees in \cite{baggingSurv} and survival random forest in \cite{randomSurvivalModel}, but also combined approaches as in \cite{dudoit}.\\ 
The standard bagging approach deploys the Efron's bootstrap (\cite{efron}), but further investigation is suggested to better understand how different resampling methods could be explored in terms of methods and applications. In fact, the use of Efron's bootstrap have one main drawback, it completely relies on the observation contained in the dataset and produces bootstrap replicas that are extremely similar one with each others. This leads to highly correlated trees that reduces the expected improvement of an ensemble approach. The use of Bayesian bootstrap methods can introduce an additional prior knowledge to the observed data, increasing the variability between datasets and improving the stability of the results and the predictive power of the model.  \\
This kind of approach already showed good results in regression and classification problems, as demonstrated in \cite{Bardis}.
In a preliminary analysis in the survival framework (see \cite{Ballante}), it emerges that the application of the Rubin bootstrap (\cite{Rubin}) and the proper Bayesian bootstrap (\cite{MuliereSecchi}) strongly improves the performances when small sample size datasets are involved.\\
The drawback observed in \cite{Ballante} is that the proposed model decreases its efficacy with the increasing of the censorship level. 
\cite{BSB} propose a proper Bayesian bootstrap for censored data based on beta-Stacy process \cite{WalkerMuliere_BSp}.\\
Based on the state of the art, our proposal explores a novel approach to integrate Bayesian bootstrap methods in bagging tree models for censored data with the main objective of improving the predictive performance, and achieving more stable and interpretable results.\\ 
Coupling Bayesian framework and censored data specific methods aims to improve the results in the most difficult cases, as working with extremely small dataset or with high percentage of censored observations, where the standard methods show a performance loss.\\

The paper is organized as follow: Section \ref{sec:methods} describes the background of the work in terms of survival setting and bootstrap methodologies, with a specific focus on proper Bayesian bootstrap algorithm in Section \ref{sec:pbb} and beta Stacy bootstrap in Section \ref{sec:bsb}. Section \ref{sec:model} contains the model proposal, Section \ref{sec:simulation} describes the computational setting and the results of the study and Section \ref{sec:conclusions} contains the conclusions of the work.
\section{Background}\label{sec:methods}
In this section, the basic setting of the survival analysis is described and the bootstrap algorithms that are applied in the proposed model are briefly presented.\\
Let be $U$ the true survival time and $C$ the censoring time. The observed data is then composed of $T = \min(U, C)$, the time until either the event occurs or the subject is censored, $\delta = I(U \leq C)$, an indicator that takes a value of 1 if the true time–to–event is observed and 0 if the subject is censored, and $\mathbf{X} = (X^1, . . . , X^p)$, a vector of $p$ numerical covariates. Data are available for $N$ independent subjects $(T_i, \delta_i, \mathbf{X}_i),\, i = 1, . . . , N$. In the rest of the paper we will us the superscript indicating the column index in the covariates matrix and the subscript for the statistical unit index. \\
The basic setup assumes that the covariate values are available at the baseline for each subject. We are also assuming that the censorship mechanism is independent from the survival times ($C$ an $U$ are independent). 
A dataset that does not contain censored observations is only a special case when $\delta_i=1, i = 1, . . . , N$. The fundamental object of survival analysis is the survival function, defined as the probability of survival beyond a certain time t: $S(t) = P(T >t)$.\\
 Survival tree models, proposed in \cite{survtrees}, are simple but powerful non-parametric predictive tools used to make inference about an unknown function $f$ that relates the survival function $S$ with a $p$ dimension vector of covariates. \\
These models are easily interpretable while still remaining competitive in terms of predictions, although they are recognized to be an unstable procedure.\\
For this reason, they are the suitable model to be put in an ensemble scheme, similarly to classification and regression framework. In this work, we focus on the bagging procedure, as defined in \cite{bagging_breiman}. \\
The basic bagging procedure consists in drawing $B$ bootstrap replicas through Efron's bootstrap algorithm (\cite{efron}) and training a decision tree for each of them, without any pruning. The final prediction is obtained aggregating the predictions of each single tree. \\ 
Bagging procedure in survival framework was proposed at first in \cite{Danneger} and \cite{Benner}, but in \cite{baggingSurv} a general method for bagging tree models was described.
A more comprehensive review on survival trees and related ensemble method can be found in \cite{review}.\\
Several different methods were explored in this area, but the opportunity to implement the Bayesian method in the bootstrap phase of the algorithm had little attention.
In \cite{Clyde}, the Rubin bootstrap was deployed in a bagging procedure, but to our knowledge, there is no counterpart for survival analysis. \\
In this work, we explore the deployment of Bayesian bootstrapping methods (proper Bayesian bootstrap and Beta-Stacy bootstrap) into the bagging procedure for survival analysis.  

\subsection{Bootstrap algorithms}
Let $\mathbf{x}=\{x_1, ..., x_n\}$ represent i.i.d. realizations from a random variable $X$, and let $\Phi$ be a functional dependent on the distribution of $X$. The definition of $X$ here is not necessary related to the covariates $\mathbf{X}$ defined in Section \ref{sec:methods}. The objective is to derive the distribution of the estimator $\hat{\Phi}$ by drawing repeated bootstrap replications from the available sample. From a Bayesian standpoint, the goal is to estimate the posterior distribution of the statistic of interest. \\ 
Considering an exchangeable sequence of real random variables $\{X_j\}$, defined on a probability space $(\Omega, F, P)$, De Finetti's representation theorem guarantees the existence of a random distribution $F$ conditionally on which the variables are i.i.d. with distribution F. The bootstrap method approximates the unknown probability distribution F with a $F^*$. Specifically, our interest lies in calculating the distribution of a statistic $\Phi(F, X)$ conditional on the sequence $X_1, ..., X_n$: $\mathcal{L}(\Phi(F,\mathbf{X})|X_1,...,X_n)$. Utilizing bootstrap methods, that distribution can be approximated as $\mathcal{L}(\Phi(F^*,\mathbf{X})|X_1,...,X_n)$, where $F^*$ is obtained through various bootstrap approaches such as Efron’s bootstrap \cite{efron}, Rubin’s bootstrap \cite{Rubin}, Proper Bayesian bootstrap \cite{MuliereSecchi} and Beta Stacy bootstrap \cite{BSB}.\\
The Efron's bootstrap was proposed in 1971 \cite{efron} and it consists in generating independently each bootstrap resample $X^*_1, ...,X^*_n$ from the empirical distribution $F_n$ of $X_1,...,X_n$. This procedure is equivalent to draw, for each bootstrap replication, a weights vector \textbf{w} for the observations $X_1,...,X_n$ from a Multinomial distribution with parameters 
$(n, \frac{1}{n}\mathbb{1}_n)$. In this way we obtain:
\begin{equation*}
F^*(x) = \sum_{i=1}^{n} \frac{w_i}{n} \mathbb{I}_{[X_i \leq x]}
\end{equation*}
where $ (w_1, ..., w_n) \sim Mult(n, \frac{1}{n}\mathbb{1}_n) $. \\
Efron's procedure assumes that the sample cumulative distribution function is the population cumulative distribution function and, under this assumption, generates a bootstrap replication $X^*_1, ...,X^*_n$ with replacement from the original sample. \\
Rubin in 1981 proposed an alternative bootstrap procedure that relies on Bayesian non-parametric framework, called Bayesian bootstrap \cite{Rubin}. Considering this procedure, the distribution $F$ is approximated by:
\begin{equation*}
F^*(x) = \sum_{i=1}^{n} w_i \mathbb{I}_{[X_i \leq x]}
\end{equation*}
where $w_i$ are randomly sampled weights distributed as 
$(w_1, ..., w_n) \sim D(\mathbb{1}_n)$. \\
The prior is assumed to be a Dirichlet process, thus the obtained posterior is again a Dirichlet process. 
In this case there is no prior distribution involved and the empirical cumulative density function $F_n$ approximates the distribution $F$, as in Efron's bootstrap resampling.\\
The main difference between the two methods is that in Rubin procedure the weights are independently drawn from a Dirichlet distribution with parameter $\alpha = 1$.
On the other hand, this approach, due to its assumptions regarding the data generating process, has the great advantage of characterizing the posterior distribution of $F^*$ given $X_1,...,X_n$,  as a Dirichlet process with parameters $nF_n$.\\
Rubin's and Efron's bootstraps are proved to be asymptotically and first order equivalent from a predictive point of view (\cite{Lo_bb}, \cite{weng}) in the sense that they estimate the conditional probability of a new observation considering only the past observations at hand.\\
From the construction of ensemble model point of view, Efron's and Rubin's bootstraps present two main drawbacks: first, they are both non informative procedures which do not take into account any prior opinions related to the distribution of $F$, second, they estimate $F$ considering only observed values included in the data at hand, giving zero probability to values not available in the sample data. Starting from these weaknesses, the aim of the present work is to introduce Bayesian bootstrap procedures, the Proper Bayesian bootstrap and the beta Stacy bootstrap, in ensemble tree models in order to improve the stability and the accuracy of the results in difficult scenarios, like small sample size data and high presence of censorship. \\
In Section \ref{sec:pbb} and \ref{sec:bsb}, a more detailed description of the Bayesian bootstrap procedures involved in the proposed model is provided.

\subsubsection{Proper Bayesian bootstrap}\label{sec:pbb}
A generalization of the Bayesian bootstrap introduced by Rubin was proposed in \cite{MuliereSecchi} (see also \cite{MuliereSecchi2003}). The name \textit{proper Bayesian bootstrap} referred to the fact that this new bootstrap algorithm allows the introduction of a prior knowledge information, represented by a distribution function $F_0$. Following the work of Ferguson in \cite{ferguson}, they define the prior of $F$ as a Dirichlet process $D(kF_0)$ where $F_0$ is a proper distribution function and $k$ represents the level of confidence in the initial choice $F_0$. The resulting posterior distribution for $F$, given a sample of $X_1,...,X_n$ from $F$, is still a Dirichlet process with parameter $(kF_0+nF_n)$, so when $k=0$ the procedure is equivalent to the Rubin's one. This bootstrap method allows to introduce explicitly prior knowledge on the data through the choice of $F_0$ and $k$. It is important to remark that, since $\phi$ is a function of F, an informative prior on F is an informative prior on $\phi$.\\
When $k>0$ it is often difficult to derive analytically the distribution of $\phi(F)$. When $F_0$ is discrete with finite support one may produce a reasonable approximation on the distribution of $\phi(F)$ by a Monte Carlo procedure obtaining i.i.d. samples from $D(kF_0)$. If $F_0$ is not discrete,  
in order to compute the posterior distribution $D(kF_0+nF_n)$, a possible way proposed in \cite{MuliereSecchi} is to first approximate the distribution $kF_0+nF_n$ through $(n+k)F^*_m$, where $F^*_m$ is the empirical distribution of an i.i.d. bootstrap resample of size $m$ generated from: 
\begin{equation}
\label{eqn:3}
G_0= \frac{k}{n+k}F_0 + \frac{n}{b+k}F_n
\end{equation}
The bootstrap resample is generated from a mixture of the empirical distribution function of $X_1,...,X_n$ and $F_0$. 
Let now define $G^*_m$ as a random distribution which, conditionally on the empirical distribution $F^*_m$ of $X^*_m$, is a Dirichlet process $D((k+n)F^*_m)$. Thus, since $G^*_m$ is given by a mixture of Dirichlet processes, using \cite{antoniak} when $m$ goes to infinite the law of $G^*_m$ weakly converges to the Dirichlet process $D(kF_0+nF_n)$. \\
The algorithm \ref{properbootstrap} shows the computational steps involved in proper Bayesian bootstrap method for the generation of $B$ bootstrap replicas of the original dataset.\\
\begin{algorithm}[ht!]
	\caption{}\label{properbootstrap}
	\vspace{0.5em}
	\textbf{Proper Bayesian bootstrap}\\
	\vspace{0.5em}
	\vspace{0.5em}
	Observations $x_1,x_2,...,x_n$\\
	for\{b in 1:B\}\{\\
		Generate m observations $x_1^*,...,x_m^*$ from $(k+n)^{-1}(kF_0 + nF_n)$\\
		Draw $w_1^b,...,w_m^b$ from $D(\frac{n+k}{m},...,\frac{n+k}{m})$;\\
		Get $\phi^b = \phi(\sum_{i=1}^mw_ix_i^*)$\}
	\vspace{0.5em}
\end{algorithm}
The conditional distribution $\mathcal{L}(\Phi(F^*,\mathbf{X})|X_1,...,X_n)$ is approximate by derive the empirical distribution function generated by $\phi_1,...,\phi_B$, where $B$ is the number of bootstrap resamples. \\

\subsubsection{Beta-Stacy bootstrap} \label{sec:bsb}
The Beta Stacy bootstrap is defined as a Bayesian non-parametric inference method that allows to proper treat censored data. It is based on the definition of Beta Stacy process provided in \cite{WalkerMuliere_BSp} and it deploys the property of the beta Stacy process of being conjugate with respect to right-censored data, allowing a simple posterior computation. The property of conjugacy to right censored observations is valid also with the beta process, but beta Stacy process allows a higher flexibility in the construction of the prior distribution.\\
Considering a beta Stacy distribution $G\sim BS(c,F)$ and $n$ possibly censored survival times $\tau_1,... \tau_n$ generated from $G$, a results from \cite{WalkerMuliere_BSp} states that the posterior distribution of $G$ conditional on $\tau_1,... \tau_n$is the beta Stacy process $BS(c^*,F^*)$, where $F^*$ and $c^*$ are the updated values of the parameters. More mathematical detail can be found in \cite{WalkerMuliere_BSp} and \cite{BSB}.  \\
The beta Stacy bootstrap approach samples from an approximation to the law of $\Phi(G^*)$, where $G^*\sim BS(c^*,F^*)$ is the posterior distribution already defined. \\
The algorithm \ref{bsboostrap} shows how to estimate $\Phi(G^*)$, as described in \cite{BSB}, that in the present case is a resampling of the posterior distribution of time-to-event data.
\begin{algorithm}[ht!]
	\caption{}\label{bsboostrap}
	\vspace{0.5em}
	\textbf{beta Stacy bootstrap}\\
	\vspace{0.5em}
 	\vspace{0.5em}
	Time-to-event data $(t_1,\delta_1),...,(t_n,\delta_n)$

 \small Sample $X_1, ..., X_m$ from $F^*$ and obtain the corresponding number D of distinct values $X_{1,m}< ... < X_{D,m}$.\\
 
Compute $\alpha_i = c^{*}(X_{i,m})\Delta F_m(X_{i,m}), \beta_i = c^{*}(X_{i,m})\bar{F}_m(X_{i,m})$ for every $i = 1, . . . , D$, where $F_m(x) =\sum_{i=1}^{m} I\{X_i \leq x\}/m$ is the empirical distribution function of $X_1, . . . , X_m$.\\

 For all $i = 1 . . . , D$, generate $U_i \sim Beta(\alpha_i, \beta_i)$ ($U_D = 1$, as $\beta_D = 0$). Let $Z_i = U_i \prod_{j=1}^{i-1}(1 - U_j)$.\\
 
 Let $G_m(x) = \sum_{i=1}^{D} I\{X_{i,m}\leq x\}Z_i$ and compute $\Phi(G_m) = f (G_mh_1, . . . , G_mh_k)$, where $G_mh_j = \sum_{i=1}^{D} h(X_{i,m})Z_i$ for all $j = 1, . . . , k$.\\
 
 Get $\Phi(G_m)$ as an approximate sample from the distribution of $\Phi(G^{*})$.
\end{algorithm}
Beta-Stacy bootstrap extends other common Bayesian bootstrap procedure. Specifically, the proper Bayesian bootstrap is a special case of Beta-Stacy bootstrap that occurs if the data are not censored and $c(x)=k$. \\
In the next session it is described how these Bayesian bootstrap approaches can be deployed in ensemble models.

\section{Generalized Bayesian ensemble survival tree (GBEST) model} \label{sec:model}
Our contributions is to develop a bagging tree model where the Efron's bootstrap method is replaced with the suitable Bayesian bootstrap methods to improve the stability and the accuracy of the results. \\
The point of interest in ensemble tree models is $\phi(F, \mathbf{X}$), a decision tree model reliant on a distribution $F$ and observed data $\mathbf{X}$. As reported in \cite{Ballante} and before in \cite{Bardis}, an approximation of the posterior distribution of $\phi$ can be derived via bootstrap procedures. Each tree is constructed by fitting the model to a bootstrap replica of the original dataset, and the resultant predictions serve as an estimate of the posterior mean.

To implement the Proper Bayesian bootstrap, we establish a prior distribution $D(k F_0)$ for $F$, which takes the form of a Dirichlet process. In the context of elucidating a response variable $y$ within the framework of a list of P covariates $x_1, ..., x_P$, the parameter $F_0$ of the Dirichlet process manifests as a joint distribution reliant on both $(\mathbf{x},y)$.

The procedure for sampling from the posterior of $\phi(F, \mathbf{X})$ is defined in \cite{Bardis} and is detailed in this document as Algorithm \ref{gbet}.

\begin{algorithm}[ht!]
	\caption{Generalized Bayesian Ensemble Tree models in survival analysis}\label{gbet}
	\vspace{0.5em}
	Training set $T$\\
	for\{b in 1:B\}\{\\
		Sample $(\mathbf{x}^*_1),...,(\mathbf{x}_m^*)$ from $(k+n)^{-1}(k F_0+n F_n)$;\\
		Draw $\mathbf{w}^b$ from $D(\frac{n+k}{m},...,\frac{n+k}{m})$;\\
            Sample $\tau^*_1,...,\tau_r^*$ from the beta-Stacy distribution, where $r$ is the number of prior sampled covariates;\\
            Match the $\tau^*_1,...,\tau_r^*$ extracted with the $\mathbf{x}^*_1,...,\mathbf{x}_m^*$ sampled from the prior considering the hazard rate estimated function;\\
		Get $\phi^b = \phi(\mathbf{w}^b)$ running weighted tree on the new sample $(\mathbf{x}^*_1,y_1^*)...,(\mathbf{x}_m^*,y_1^*)$	\}
	\vspace{0.5em}
\end{algorithm}

The bootstrap resample $\mathbf{x}^*_1, , ..., \mathbf{x}^*_m$ is created through a mixture of distributions involving the prior estimate $F_0$ and the empirical distribution $F_n$. Given the assumption of covariate independence, when a new observation in the bootstrap resample is generated from $F_0$, a new vector of covariates $\mathbf{x}$ is formed based on the original prior distributions $F_0(x_k)$.
Independently, a vector of time-to-event data $\tau_1,...,\tau_r$ is generated through the beta-Stacy bootstrap. The linkage of these two vectors is to assign each time to the vector of covariates with the higher value of the estimated hazard rate.\\
On these datasets, the $B$ survival trees are trained. The output of the ensemble model is a bootstrap-aggregated version of the estimated conditional survival function $S$ for a new observation $\mathbf{x}_{new}$, calculated by aggregating all training observations falling into the same leaves as $\textbf{x}_{new}$. This is expressed as $\hat{S}^B_A(\cdot\vert\mathbf{x}_{new}) = \hat{S}_{L^B_A(\mathbf{x}_{new})}(\cdot)$, as suggested in \cite{baggingSurv}.

The primary novelty compared to traditional ensemble procedures is that the prior $F_0$ enables the generation of new observations not present in the training set, which can enhance our predictive model.\\
\subsection{Comparison with the previous proposal}
A similar idea was already explored in \cite{Ballante}, where the survival times of the prior generated data where assigned using a parametric predictive model. This approach presented a drawback especially when the censorship level of the dataset increases or when the dataset is really small. \\
In this work, we propose an improved version of this model that is more robust in these difficult settings and the different performances will be shown.\\
Even if the main structure is really similar between the two models, the new one introduces the exploitation of Beta Stacy bootstrap method to overcome the instability of the older model. In fact, the observations that are drawn from the prior distribution come without a time-to-event label. In the older model, a Weibull regression was proposed to predict these labels, but this parametric model showed greater instability in the prediction, providing unreasonable results. In the most severe cases, this can compromise the contribution of the prior knowledge, which is the main point of the proposal. \\
For the reasons described, we believe that the proposed model can provide more stable results and, in some scenarios, outperform both the classical models and the previous version of the same model.\\
\section{Empirical results on simulated data}\label{sec:simulation}
The analysis is performed in R 4.2.0. 
Fifty datasets are simulated for each setting, with each dataset randomly split into 50\% for training and 50\% for the test. The errors obtained on the test sets are reported as averages and non parametric confidence intervals.\\
The simulated datasets consist of 5 numerical covariates sampled from a uniform distribution $U(0,10)$ and a time-to-event target variable. The simulation of the survival times is based on a Weibull survival regression. 
The censorship indicator is set when the survival time is higher than the maximum observation time allowed.\\
For the proper Bayesian bootstrap, the prior functions are uniform distributed on the range of each covariate in the original dataset, and the parameter $k$ is set such that the weight $w=\frac{k}{k+n}$ assigned to the prior $F_0$ is equal to 0, 0.1 and 0.2. The case $w=0$ corresponds to the Rubin bootstrap method. \\
The proposed model is compared both with the most common models in survival analysis: the Survival Random Forest (\cite{randomSurvivalModel}) and the Cox model (\cite{Cox}), as well as with the previous version of the model defined in \cite{Ballante}.
In the sections related to the application to simulated and real data, Section \ref{sec:simulation} and Section \ref{sec:real_data}, the proposed model is referred to as GBEST\_BSB and the previous version as GBEST.\\
The comparison are made by varying the sample size or the censorship level of the dataset.\\
The predictive performances of the different models are investigated on simulated data and evaluated in terms of Integrated Brier Score (IBS). Consider the difficulties of prediction in the right tail of time distribution, in order to reduce the variability the Integrated Brier Score is evaluated in $[0, 0.8*T_max]$.
The models are visually compared in terms of means and non-parametric confidence intervals (quantiles of level 0.05 and 0.95). Detailed results in terms of descriptive statistics (median and quartile) and statistical comparisons are also reported. 

\subsection{Comparison with the literature models} \label{literature}
\begin{figure}[ht!]
	\centering
	\begin{subfigure}[b]{\linewidth}
        \centering
		\includegraphics[width=0.6\linewidth]{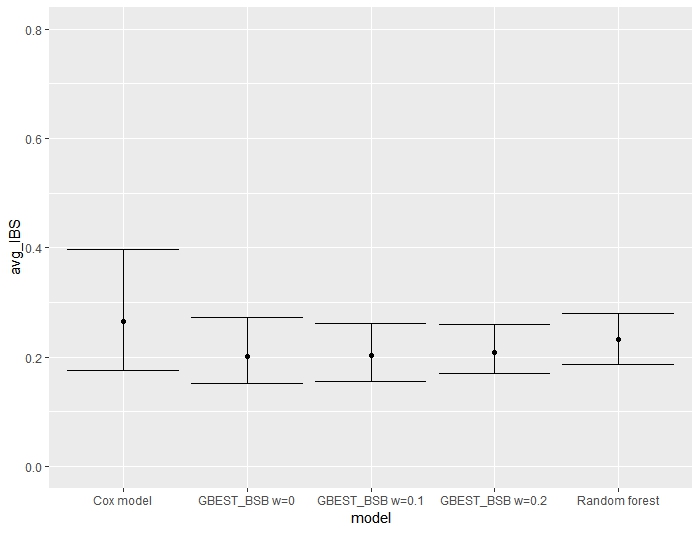}
		\caption{N=50. Censorship level: 10\%.}
	\end{subfigure}
	\begin{subfigure}[b]{\linewidth}
     \centering
		\includegraphics[width=0.6\linewidth]{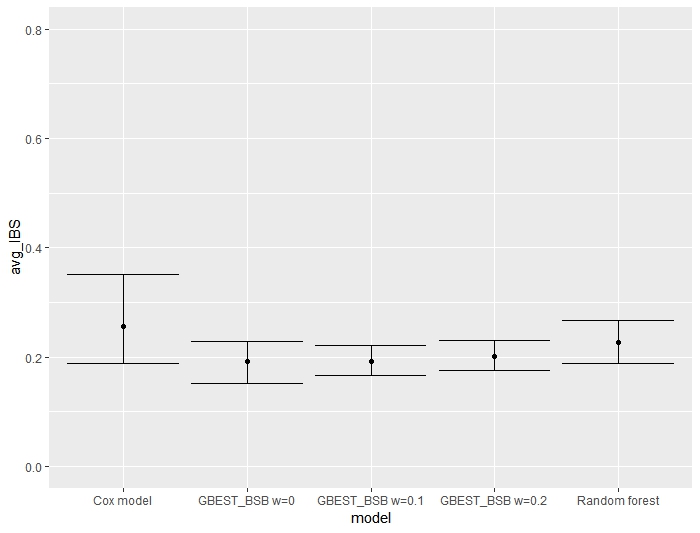}
		\caption{N=75. Censorship level: 10\%.}
	\end{subfigure}
	\begin{subfigure}[b]{\linewidth}
     \centering
		\includegraphics[width=0.6\linewidth]{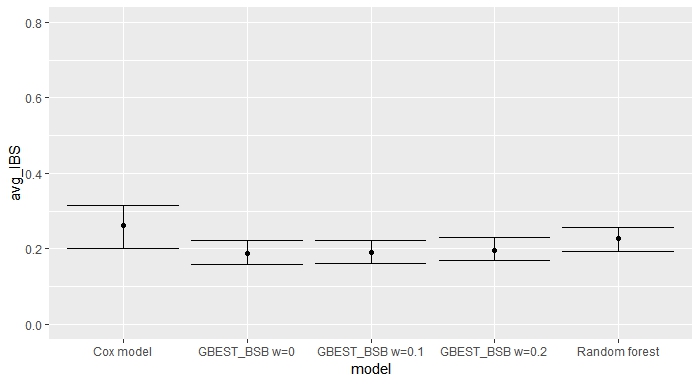}
		\caption{N=100. Censorship level: 10\%.}
        \end{subfigure}
	 \caption{Comparison of mean and non-parametric confidence intervals for IBS obtained for the 50 simulated datasets at a censorship level of 10\%.. The models are the Cox model, the proposed model with different weights (GBEST\_BSB w=0, GBEST\_BSB w=0.1, GBEST\_BSB w=0.2) and the Random forest, respectively.} 
  \label{fig:samplesize_10}
\end{figure}

\begin{figure}[ht!]
	\centering
	\begin{subfigure}[b]{\linewidth}
     \centering
		\includegraphics[width=0.6\linewidth]{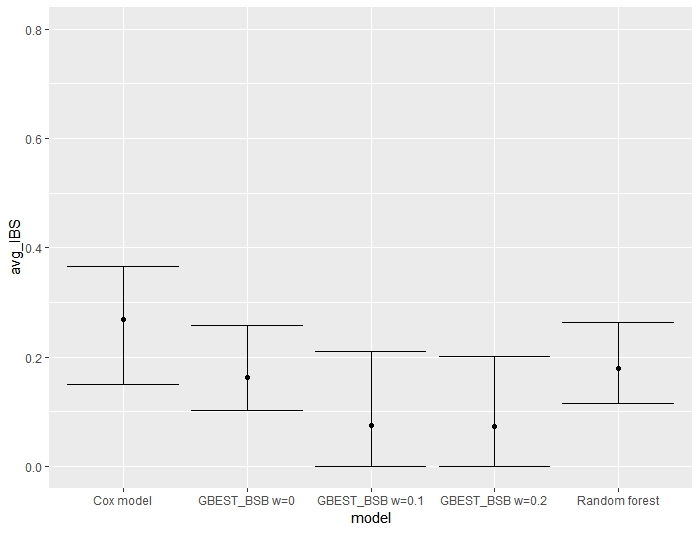}
		\caption{N=50. Censorship level: 70\%.}
	\end{subfigure}
	\begin{subfigure}[b]{\linewidth}
     \centering
		\includegraphics[width=0.6\linewidth]{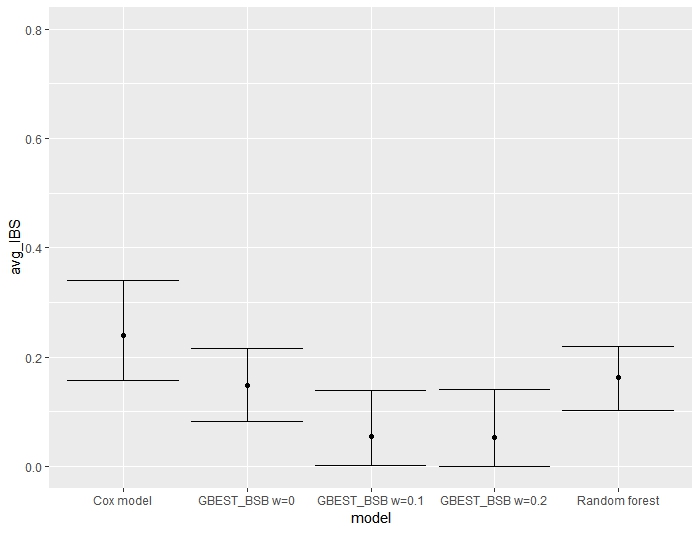}
		\caption{N=75. Censorship level: 70\%.}
	\end{subfigure}
	\begin{subfigure}[b]{\linewidth}
     \centering
		\includegraphics[width=0.6\linewidth]{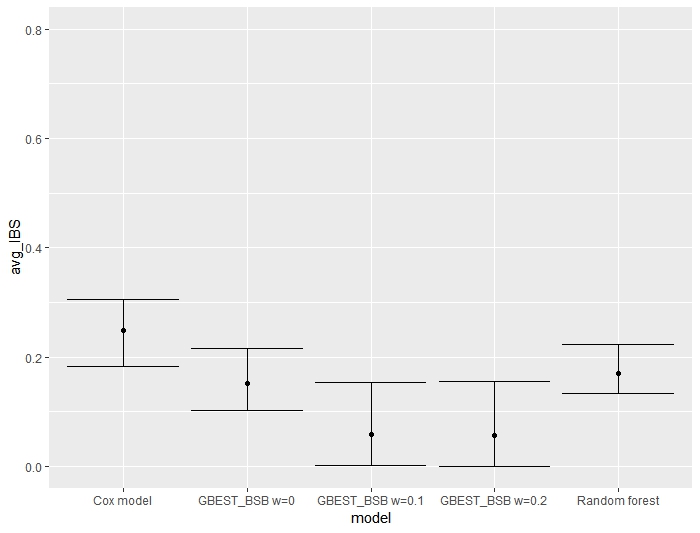}
		\caption{N=100. Censorship level: 70\%.}
        \end{subfigure}
	 \caption{Comparison of mean and non-parametric confidence intervals for IBS obtained for the 50 simulated datasets at a censorship level of 70\%. The models are the Cox model, the proposed model with different weights (GBEST\_BSB w=0, GBEST\_BSB w=0.1, GBEST\_BSB w=0.2) and the Random forest, respectively.} 
  \label{fig:samplesize_70}
\end{figure}

The first aspect investigated is how the sample size and censorship levels influences the performance of the models considered compared to the literature models. The sample sizes considered are N=50 (N=25 for the training and N=25 for the test), N=75 (N=38 for the training and N=37 for the test) and N=100 (N=50 for the training and N=50 for the test). The censorship rate is set at 10\% and 70\% of the sample size, in order to consider two different scenarios in terms of censorship level. \\ 
Figure \ref{fig:samplesize_10} and \ref{fig:samplesize_70} show that the proposed model has lower levels of error and smaller confidence intervals compared to the Cox model and Random Forest in each setting. The difference is more evident with respect to the Cox model than to Random Forest, that especially for low levels of censorship seems almost equivalent to the proposed model.The different weights assigned to the prior have a small influence on the results when the censorship level is low but contribute of the  the prior becomes relevant for every sample size in a setting with high level of censorship. 

\subsection{Comparison with the previous proposal}
\begin{figure}[ht!]
	\centering
	\begin{subfigure}[b]{\linewidth}
    \centering
		\includegraphics[width=0.6\linewidth]{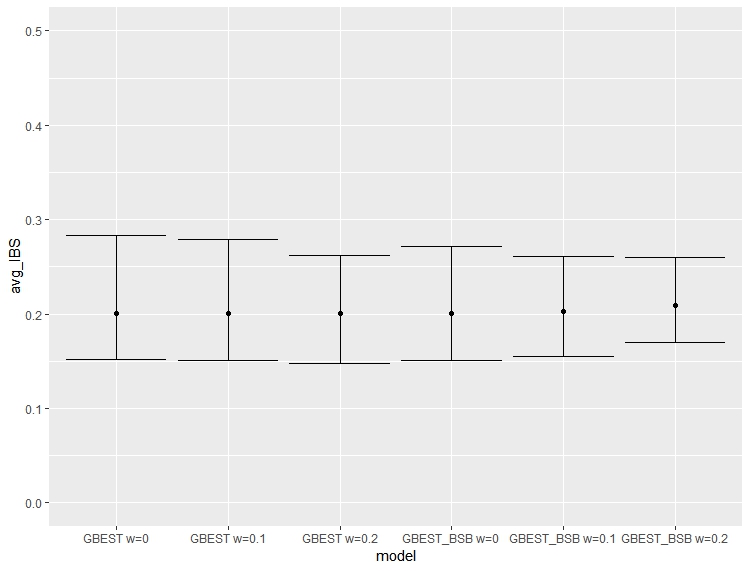}
		\caption{N=50. Censorship: 10\%.}
	\end{subfigure}
	\begin{subfigure}[b]{\linewidth}
    \centering
		\includegraphics[width=0.6\linewidth]{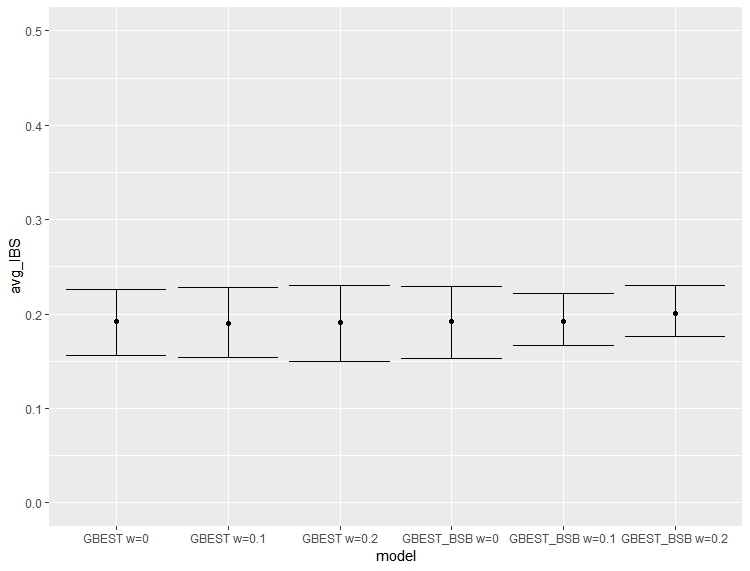}
		\caption{N=75. Censorship: 10\%.}
	\end{subfigure}
 \begin{subfigure}[b]{\linewidth}
 \centering
		\includegraphics[width=0.6\linewidth]{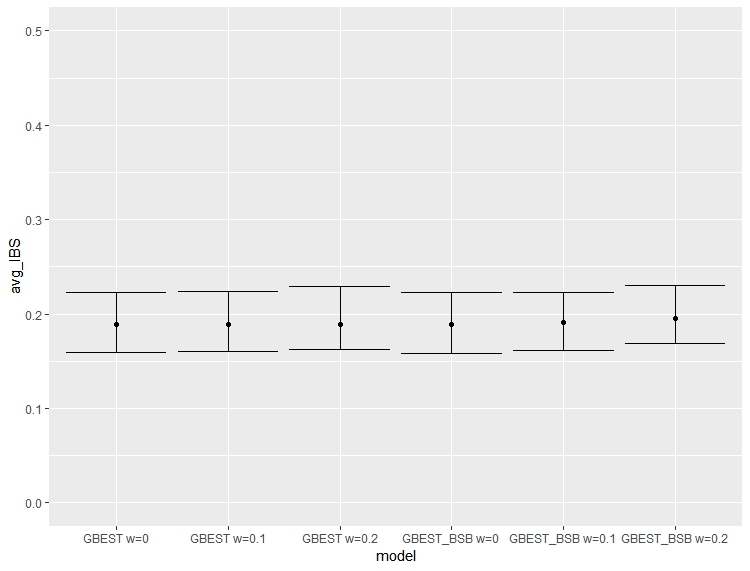}
		\caption{N=100. Censorship: 10\%.}
	\end{subfigure}
	\caption{Comparison of mean and nonparametric confidence intervals for IBS obtained for the 100 simulated datasets. The models are the old model with different weights (GBEST w=0, GBEST w=0.25, GBEST w=0.50) and the proposed model with different weights (GBEST\_BSB w=0, GBEST\_BSB w=0.1, GBEST\_BSB w=0.2), respectively.}
 \label{fig:samplesize_old10} 
\end{figure}

\begin{figure}[ht!]
	\centering
	\begin{subfigure}[a]{\linewidth}
        \centering
		\includegraphics[width=0.6\linewidth]{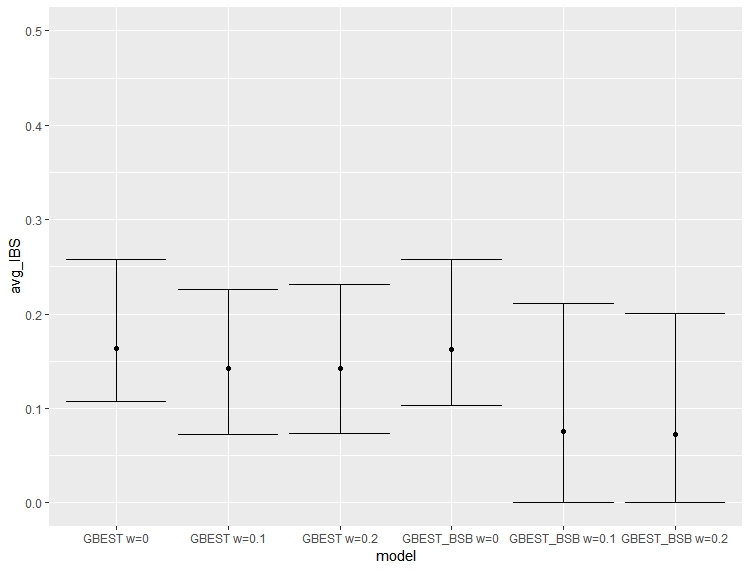}
		\caption{N=50. Censorship: 70\%.}
	\end{subfigure}
	\begin{subfigure}[b]{\linewidth}
        \centering
		\includegraphics[width=0.6\linewidth]{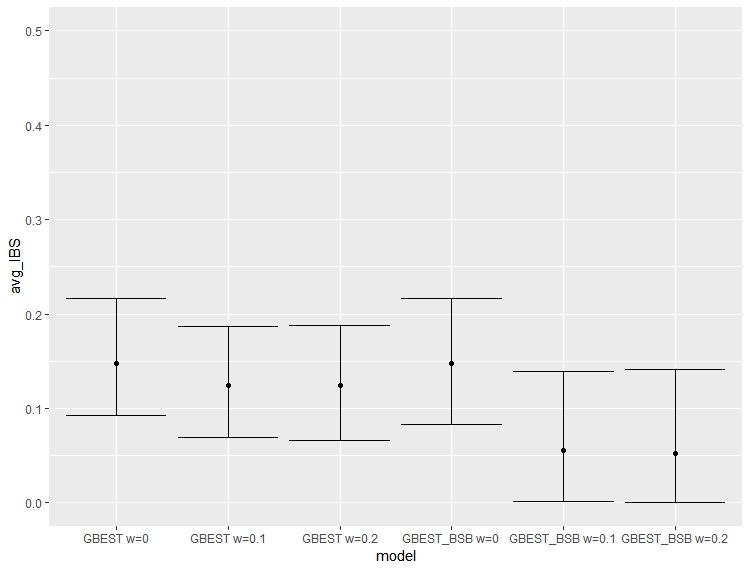}
		\caption{N=75. Censorship: 70\%}
	\end{subfigure}
 \begin{subfigure}[c]{\linewidth}
        \centering
		\includegraphics[width=0.6\linewidth]{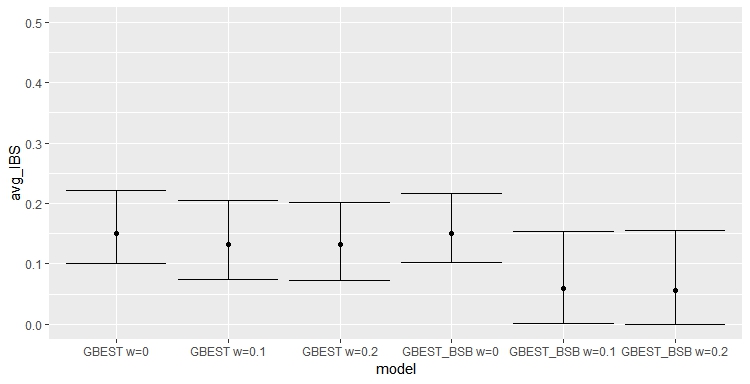}
		\caption{N=100. Censorship: 70\%.}
	\end{subfigure}
	\caption{Comparison of mean and nonparametric confidence intervals for IBS obtained for the 100 simulated datasets. The models are the old model with different weights (GBEST w=0, GBEST w=0.25, GBEST w=0.50) and the proposed model with different weights (GBEST\_BSB w=0, GBEST\_BSB w=0.1, GBEST\_BSB w=0.2), respectively.}
 \label{fig:samplesize_old70} 
\end{figure}

Comparing the results of the new method with the previous version, Figure \ref{fig:samplesize_old10} shows that the performances do not differ significantly when the censorship level is low.\\
Instead, Figure \ref{fig:samplesize_old70} shows that, in presence of a high level of censorship, the new version of the method shows a relevant improvement when the prior is included.\\

In conclusion, simulation results underline that the proposed model introduces substantial predictive improvements in terms of IBS with respect to the most common models in the literature and its previous version. Its application is relevant when the censorship level is high, even in case of small and moderate sample sizes. 
The weights assigned to the prior (w=0, w=0.1 and w=0.2) are chosen after an extensive simulation that shows that an higher value (as a rule of thumb w$\geq$0.3) introduce higher level of errors and variability.\\
\subsection{Final results}
In order to confirm the results obtained, a linear regression analysis of the Integrated Brier score has been performed. Fifty datasets are generated for every combination of the parameters included in the model.\\
The covariate of the model are:
\begin{itemize}
    \item N. The sample size of the simulated datasets, that corresponds to the double of the training set. The values N=50, 75, 100, 200 are included.\\
    \item cens. The level of censorship of the datasets. 10\%, 40\% and 70\% are included.\\
    \item model. Different models are compared. The reference level is the proposed model with a weight $w=0.1$, compared to the old version of the model with the same weight (GBEST $w=0.1$), to the Random Forest and Cox model.\\
    \item prior. Two different choices of prior distribution are compared, the uniform prior (unif) against a normally distributed prior that is the reference level. \\
    \item covariate. The distribution of the simulated data are included in the model. Two different settings are considered, variables independently sampled by a uniform distribution U(-1,1) and five variables independently sampled by a normal distribution N(0,1).\\
\end{itemize}
The logit transformation is applied to the IBS in order to transform the range $[0,1]$ to $(-\infty,+\infty)$. Since the variable still doesn't satisfy the normality assumption, and other common transformations do not solve the problem, a bootstrap procedure is applied to obtain coefficients, standard deviation and 95\% non parametric confidence intervals.
In Table \ref{tab:lm} the results of the linear regression model are shown.

\begin{table}[ht]
\centering
\begin{tabular}{rrrrr}
  \hline
 & coefficient & standard deviation & lower bound CI & upper bound CI \\ 
  \hline
(Intercept) & -1.5128 & 0.0207 & -1.5496 & -1.4704 \\ 
  N & -0.0002 & 0.0001 & -0.0004 & -0.0001 \\ 
  cens & -0.8140 & 0.0229 & -0.8592 & -0.7737 \\ 
  model GBEST w=0.1 & \textbf{0.2622} & 0.0203 & 0.2219 & 0.2987 \\ 
  model Cox model & \textbf{0.6397} & 0.0212 & 0.5998 & 0.6776 \\ 
  model Random forest & \textbf{0.4556} & 0.0203 & 0.4197 & 0.4956 \\ 
  prior unif & -0.0228 & 0.0095 & -0.0403 & -0.0029 \\ 
  covariate unif & 0.0055 & 0.0102 & -0.0130 & 0.0253 \\ 
   \hline
\end{tabular}
\caption{Linear regression model for the logit transformation of the Integrated Brier score with a bootstrap procedure.}
 \label{tab:lm}
\end{table}

The only non-significant variable is the distribution of the covariates, that in this setting seems to have no influence on the results. The sample size have a limited but negative impact, which means that an increase of the sample size leads to slightly better results. Also the censorship level seems to have a negative impact in this setting. \\
The conclusive results are shown in terms of comparison of the models considered. The three models showed in the linear model results have significantly positive coefficient that corresponds to higher level of error with respect to the proposed one.

\section{Real data analysis}\label{sec:real_data}
The performance of the proposed model is demonstrated using a real dataset on bladder cancer recurrences, sourced from the survival package in R (\cite{survival-package}). The selected version  of the dataset comprises 85 subjects with nonzero follow-up, who were assigned to either thiotepa treatment or a placebo. Only the first recurrence is selected for any patient, as the recurrent event feature falls outside the scope of this work. The status variable indicates recurrence with 1 and 0 for everything else, with the corresponding times provided. Other predictors in the dataset, besides treatment, include the initial number of tumors and the size of the largest initial tumor.\\
The censorship level of the dataset is the 45\%, with 38 censored subjects and 47 events.\\
The results are presented as the average IBS and parametric confidence intervals obtained from a 5-fold cross validation exercise.\\
Regarding the proper Bayesian bootstrap parameters, the priors are uniform distributions covering the range
of each covariate in the original dataset, and the parameter $k$ is set such that the
weight $w = \frac{k}{k+n}$ assigned to the prior $F_0$ is equal to 0, 0.25 and 0.5.
The proposed model is compared with the most common models in survival analysis: the Survival Random Forest (\cite{randomSurvivalModel}) and the Cox model (\cite{Cox}).

\begin{figure}[ht!]
    \centering \includegraphics[width=\linewidth]{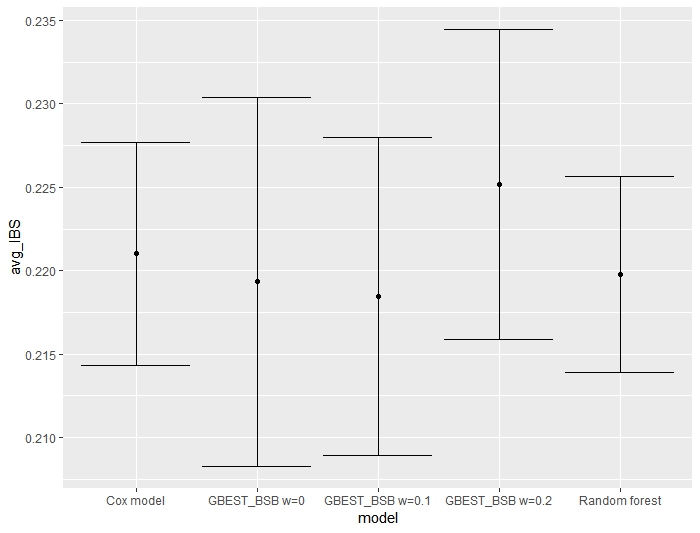}
    \caption{Comparison of mean and parametric confidence intervals for IBS obtained in a 5-fold cross validation exercise. The models are the Cox model, the proposed model with different weights (GBEST\_BSB w=0, GBEST\_BSB w=0.1, GBEST\_BSB w=0.2) and the Random forest, respectively.}
    \label{fig:res_real}
\end{figure}

\begin{table}[ht!]
\centering
\begin{tabular}{lrr}
  \hline
  model & avg\_IBS & sd\_IBS \\ 
  \hline
    GBEST\_BSB w=0 & 0.219 & 0.013 \\ 
   GBEST\_BSB w=0.1 & \textbf{0.218} & \textbf{0.011} \\ 
   GBEST\_BSB w=0.2 & 0.225 & 0.011 \\ 
  Random forest & 0.220 & 0.007 \\ 
 Cox model & 0.221 & 0.008 \\ 
   \hline
\end{tabular}
\caption{Results from various models applied to the real dataset were evaluated in terms of IBS. Presented are the averaged values and standard deviations of the 5-fold cross validation analysis are presented.} \label{tab:res_real}
\end{table}

Results reported in Figure \ref{fig:res_real} and in Table \ref{tab:res_real} show that the proposed model with a prior weight of 0.1 obtain the best results with an average IBS of 0.218, while Random forest obtain an average IBS of 0.220.. \\

\section{Conclusion}\label{sec:conclusions}
This paper introduces a novel bagging tree model (GBEST), utilizing the Proper Bayesian Bootstrap coupled to the Beta Stacy bootstrap. The results demonstrate its superior potential compared to traditional survival models when high levels of censorship and small sample size data are involved. Our proposed model exhibits increased stability and competitive or better performance in a simulated environment.\\
The inclusion of synthetic data generated from a prior distribution, which were not originally part of the dataset, enhances the stability of the final ensemble model, particularly in scenarios involving small sample sizes and high levels of censorship.\\
Given the potential of addressing small sample size and high censorship level issues across various domains, we believe that the application of our proposed method holds promise for enhancing the reliability of results in real-world problem-solving.\\
Future developments will involve extending the model to handle categorical variables and addressing the covariance structures within the analyzed data.

\backmatter

\section*{Declarations}
The authors report there are no competing interests to declare.
No funding was received for conducting this study.\\
The R code related to the proposed model will be available at \url{https://github.com/eballante}.

\section*{Author contributions}
All authors contributed to the study conception and design. The design of the computational aspects and the code writing were performed by EB. The first draft of the manuscript was written by EB and all authors commented on previous versions of the manuscript. All authors read and approved the final manuscript.

\bibliography{sn-bibliography}

\end{document}